\documentclass[%
 reprint,
 amsmath,amssymb,
 aps,
showkeys
]{revtex4-2}

\usepackage{graphicx}
\usepackage{dcolumn}
\usepackage{bm}
\usepackage{textgreek} 
\begin{document}

\preprint{Testa-Adhesion}

\title{Switchable Adhesion of Soft Composites Induced by a Magnetic Field}

\author{Paolo Testa$^{1,2.3}$}
\email{paolo.testa@mat.ethz.ch}
\author{Beno\^{i}t Chappuis$^{3}$}
\author{Sabrina Kistler$^{3}$}
\author{Robert W. Style$^{3}$}
\author{Laura J. Heyderman$^{1,2}$}
\email{laura.heyderman@mat.ethz.ch}
\author{Eric R. Dufresne$^{3}$}%
\email{eric.dufresne@mat.ethz.ch}

\affiliation{
$^{1}$ Laboratory for Mesoscopic Systems, Department of Materials, 
ETH Z\"urich, 8093 Z\"urich, Switzerland \\
$^{2}$Paul Scherrer Institute,
5232 Villigen, Switzerland
$^{3}$ Laboratory of Soft and Living Materials, Department of Materials, 
ETH Z\"urich, 8093 Z\"urich, Switzerland 
}

\date{\today}

\begin{abstract}
Switchable adhesives have the potential to improve the manufacturing and recycling of parts and to enable new modes of motility for soft robots.
Here, we demonstrate magnetically-switchable adhesion of a two-phase composite to non-magnetic objects. 
The composite's continuous phase is a silicone elastomer, and the dispersed phase is a magneto-rheological fluid.
The composite is simple to prepare, and to mould to different shapes.
When a magnetic field is applied, the magneto-rheological fluid develops a yield stress, which dramatically enhances the composite's adhesive properties.
We demonstrate up to a nine-fold increase of the pull-off force of non-magnetic objects in the presence of a 250 mT field.
\end{abstract}

\maketitle

Adhesion based on chemical bonding and cross-linking, as found in epoxies, is irreversible. 
In contrast, adhesives based on physical interactions,  such as commercial pressure-sensitive adhesives (PSAs) \cite{Abbott2015, Gay1999}, display reversible adhesion and can even be re-usable.
To facilitate removal, it is desirable to be able to switch an adhesive  between  strong and weak adhesive states, enabling a host of novel applications \cite{Meitl2006, Kim2008, Croll2019, Song2019, Ye2016}.
\begin{figure}
    \centering
    \includegraphics{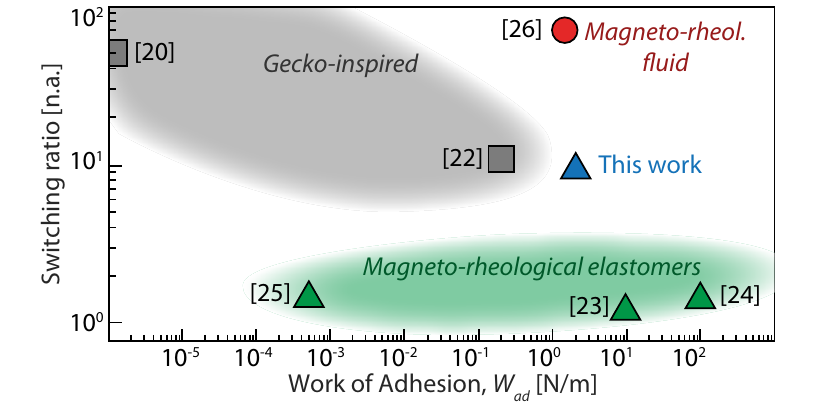}
    \caption{A comparison of the performance of the composite used in this work with other magnetically switchable adhesives found in literature. All points are for solid adhesives, except the red circle which is a layer of a magneto-rheological fluid.
    Adapted from \cite{Croll2019}.}
    \label{fig:ashby-plot}
\end{figure}
Thus far, this has been achieved with a variety of trigger mechanisms.
For example, heat can actuate hot-melt adhesives \cite{Derail1997,Li2008}, shape-memory polymers \cite{Eisenhaure2014, Eisenhaure2013}, and liquid crystal elastomers \cite{Ohzono2019}, while humidity can actuate hydrogel-based adhesives \cite{Xue2013, Cho2019}.
Mechanically-switchable adhesives can be achieved by using gecko-inspired, patterned interfaces with a directional response \cite{Minsky2017, Boesel2010a,Arzt2003, Schubert2007, Sitti2003}.
These work well, but have the disadvantages of either being slow to trigger (in the case of adhesives relying on the diffusion of heat or humidity), or relying on complex lithographic techniques (for gecko-inspired adhesives).

A promising alternative is the use of magnetic fields to switch adhesives.
These can be triggered instantaneously by the use of electromagnets, and thus offer fast switchability \cite{Croll2019}.
However, work is still needed to optimise their performance.
Shown in Fig. \ref{fig:ashby-plot} are the previously reported adhesives in terms of their work of adhesion, $W_{ad}$, and switching ratio (the difference in maximum adhesive force between `on' and `off' states).
Broadly speaking, the results fall into two groups.
Gecko-inspired magnetic adhesives \cite{Northen2008a, Gillies2013,Drotlef2014b}, exploit magnetic fields to actuate fibrillar structures on the surface of the material. 
They have high switching ratios, but at the expense of lower $W_{ad}$.
Dry, magneto-rheological elastomers utilize a change in mechanical properties to achieve switchable adhesion, and have higher $W_{ad}$, but lower switching ratio \cite{Krahn2015, Risan2015,Pang2020}.

Ideally, we would like to combine high switching ratio and high $W_{ad}$ in a moldable solid adhesive.
A suggestion of how to achieve this comes from previous work using a layer of a magneto-rheological fluid (MRF) as an adhesive layer \cite{Ewoldt2011,Watanabe2013}.
This field-activated fluid features an excellent switching ratio and a moderate to high work of adhesion, as shown by the red point in Fig. \ref{fig:ashby-plot}.
Despite excellent performance by theses metrics, its use is limited by its fluid nature.
It cannot be used to mold a part of a fixed shape, and, used as an adhesive film, it is hard to maintain intact after repeated adhesion.

\begin{figure*}
\includegraphics[width=\textwidth]{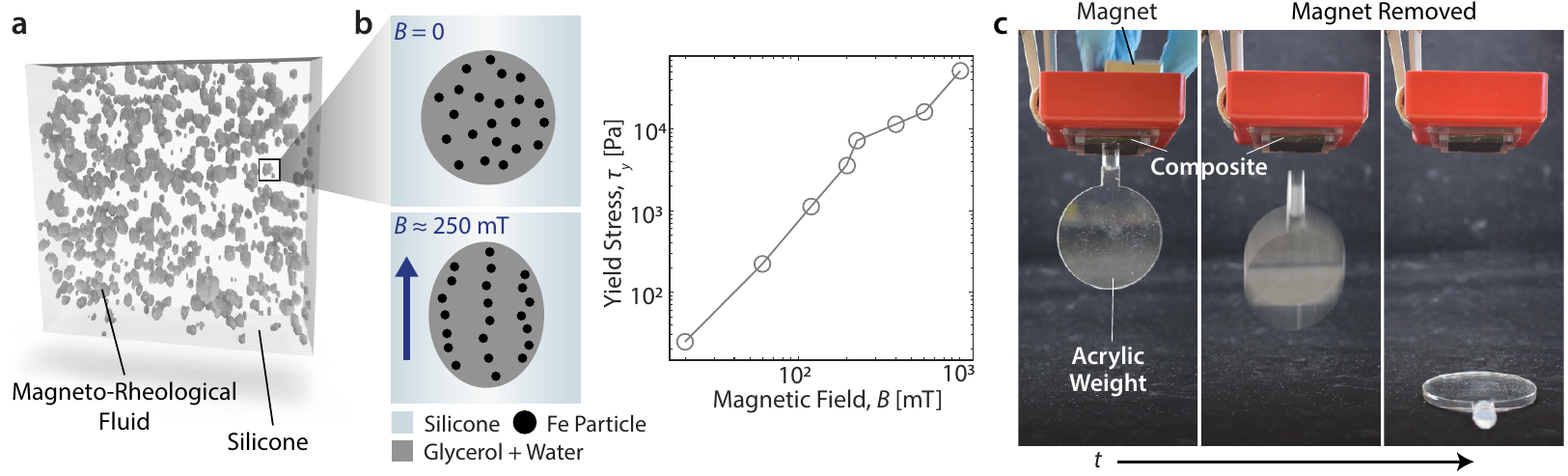}
\caption{\label{fig:overview} 
Magnetically switchable adhesion in a magneto-rheological fluid (MRF) -- silicone composite.
a) X-ray tomogram showing droplets of MRF dispersed in a silicone elastomer matrix.
b) \emph{left} Schematic illustration of the alignment of magnetic particles within a droplet under application of an external magnetic field.   \emph{right} Measured yield stress of the MRF with applied magnetic field.
c) Demonstration of magnetically-switchable adhesion.  Here, a weight made of non-magnetic acrylic sticks to the composite when a magnetic field is applied, but detaches under its own weight when the magnetic field is removed, see Movie S1.}
\end{figure*}

Here, we  combine the adhesive potential of MRF's with the stability and flexibility of a soft silicone elastomer, by simply dispersing MRF droplets in a silicone matrix (Fig. \ref{fig:overview}) \cite{Testa2019}.
With the field off, the resulting material is  soft enough to conform to a rough surface \cite{Dahlquist1969}.
When a magnetic field is applied, the MRF inclusions develop a yield stress \cite{Laun2007}, shown in Fig. \ref{fig:overview}b.
This dramatically changes  its adhesive properties, as illustrated in Fig. \ref{fig:overview}c and Movie S1, where a piece of acrylic plastic adheres to the composite when the magnetic field is on, then  falls off when the  magnetic field is removed.  
In this case, the object can stay adhered to the composite for several hours, and detaches within  a few seconds after the magnetic field is removed.

The composite is fabricated by dispersing MRF droplets into liquid silicone at a volume fraction of 30\% (\emph{c.f.} Materials \& Methods). 
After cross-linking the silicone, the  droplets are trapped in an elastic matrix (Fig. \ref{fig:overview}a) with Young's modulus tuned from roughly 4 to 40 kPa.
The MRF itself is a dispersion of micron-sized, carbonyl-iron particles (80\% by weight) in a 50:50 mixture of water and glycerol (see further details in the Materials \& Methods). 

\begin{figure*}
\centering
\includegraphics[width=\textwidth]{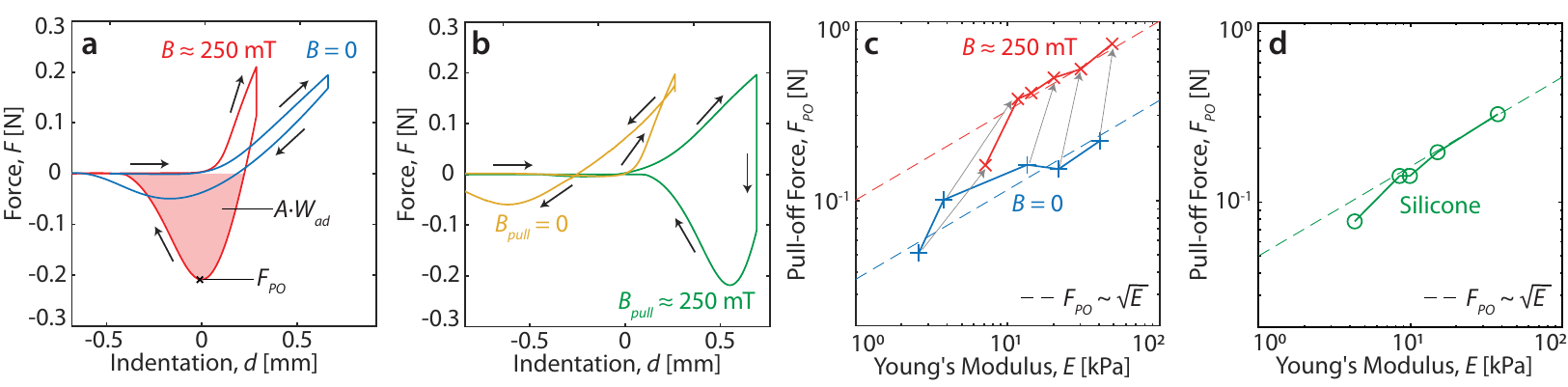}
\caption{ \label{fig:mech} 
Influence of the magnetic field on adhesion.
a) Typical force-displacement curves for a $4~\mathrm{kPa}$ composite with (red curve) and without (blue curve) an applied magnetic field of $\approx 250~\mathrm{mT}$.
b) The same material indented with the magnetic field  applied only during pull-off (green curve), or only during indentation (yellow curve). 
c)  Pull-off force versus composite modulus, with (red data) and without (blue data) an applied  magnetic field. Arrows connect data-points from the same sample.  d) Pull-off force versus Young's modulus for pure silicone (green data).  In c,d) the pull-off force increases approximately as $\sqrt{E}$, as indicated by the dashed lines.
In all panels the indentation was performed at 1 mm/min. In the last two panels, the indentation depth was 0.75 mm.}  
\end{figure*}

We quantified the adhesion of this material with a probe-tack test, as described in the Materials \& Methods. 
Briefly, we  indented a $h=3$ mm thick sample with a rigid plastic cylindrical indenter of radius $a=4.8$ mm.
By using a plastic indenter, we were able to accurately characterise the adhesion with and without a magnetic field.
Analysis of the resulting force-extension curves provided the pull-off force, $F_{PO}$, effective work of adhesion, $W_{ad}$ (both defined in Fig. \ref{fig:mech}a), and  composite elastic modulus, $E$.

Typical force-indentation results are shown in Fig. \ref{fig:mech}a. 
The results with no magnetic field  are given by the blue curve, and the response when $B\approx$ 250 mT is shown by the red curve.
These results clearly demonstrate a field-induced increase in $F_{PO}$, $W_{ad}$, and $E$ (the latter is calculated from the slope at small indentations, and its dependence on $B$ is discussed in \cite{Testa2019}). 
Significantly, $F_{PO}$ and $W_{ad}$ do not depend on whether the magnetic field was on or off during loading.
When the magnetic field is  applied only after loading, $F_{PO}$ and $W_{ad}$ are still increased, as shown by the green curve in Fig. \ref{fig:mech}b.
When the magnetic field is removed after loading (yellow curve), $F_{PO}$ and $W_{ad}$ are similar to the $B=0$ case in Fig. \ref{fig:mech}a.

For a fixed cylindrical geometry, the pull-off force from a linear elastic substrate should be proportional to $\sqrt{W_{ad}E}$ \cite{Shull1997}.  
To test whether the field-driven increase of the composite modulus is sufficient to capture the observed enhancement of adhesion, we prepared magnetic composites with a range of silicone Young's moduli between $\approx4$ and $\approx40$ kPa.
We additionally prepared a control set of pure silicone samples within the same Young's modulus range. 
For each composite, we measured the pull-off force and stiffness at a speed of 300 \textmugreek m/min, with and without the magnetic field. 
At a fixed magnetic field, the pull-off force increased with stiffness for both the composite  (Fig. \ref{fig:mech}c) and pure silicone samples (Fig. \ref{fig:mech}d), as expected. 
However, for the composite at the same stiffness, the pull-off force is higher when the magnetic field is on (Fig. \ref{fig:mech}c). 
Thus, a field-induced increase of the elastic modulus cannot  explain the observed field-enhanced adhesion.

\begin{figure}
\includegraphics{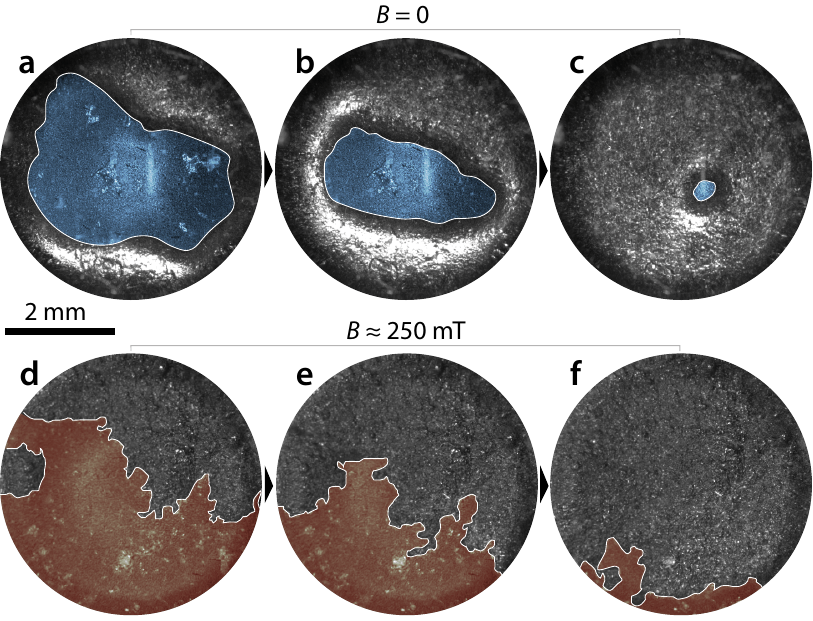}
\caption{\label{fig:crack_front}
Visualization of the delamination front.
Sequential images of the interface between a transparent indenter and composite material whose elastic modulus is $\approx 4$ kPa  when $B = 0$ during pull-off. 
The retracting delamination front is traced in white, and the blue and red false color indicates regions in contact with the indenter.
a) - c) Delamination with no applied magnetic field.
d) - f) Delamination with an $\approx 250$ mT applied field.}
\end{figure}

We hypothesize that the increased pull-off force is due to enhanced dissipation within the MRF droplets, which toughens the composite, and hinders propagation of the interfacial cracks which underlie adhesive failure.
As shown in Fig. \ref{fig:crack_front}, we observed the adhesive interface during retraction, by imaging through a fixed acrylic indenter of radius 4.8 mm (see Materials \& Methods).
For the composite without an applied magnetic field ($E \approx$ 4 kPa), we find that the delamination front has a smooth shape, similar to the delamination of pure silicone (Materials \& Methods).
When a magnetic field of $B \approx 250$ mT is applied to the composite, the modulus of the composite increases to $E \approx$ 12 kPa, and the  delamination front is very rough, similar to the delamination from an MRF with $B>0$, as shown in  supplemental Fig. \ref{fig:front-all}.
This roughening of the crack front is typical of delamination of materials that have a significant dissipative component \cite{Creton2003,Yuk2016, Crosby2000}.

\begin{figure*}
\centering
\includegraphics{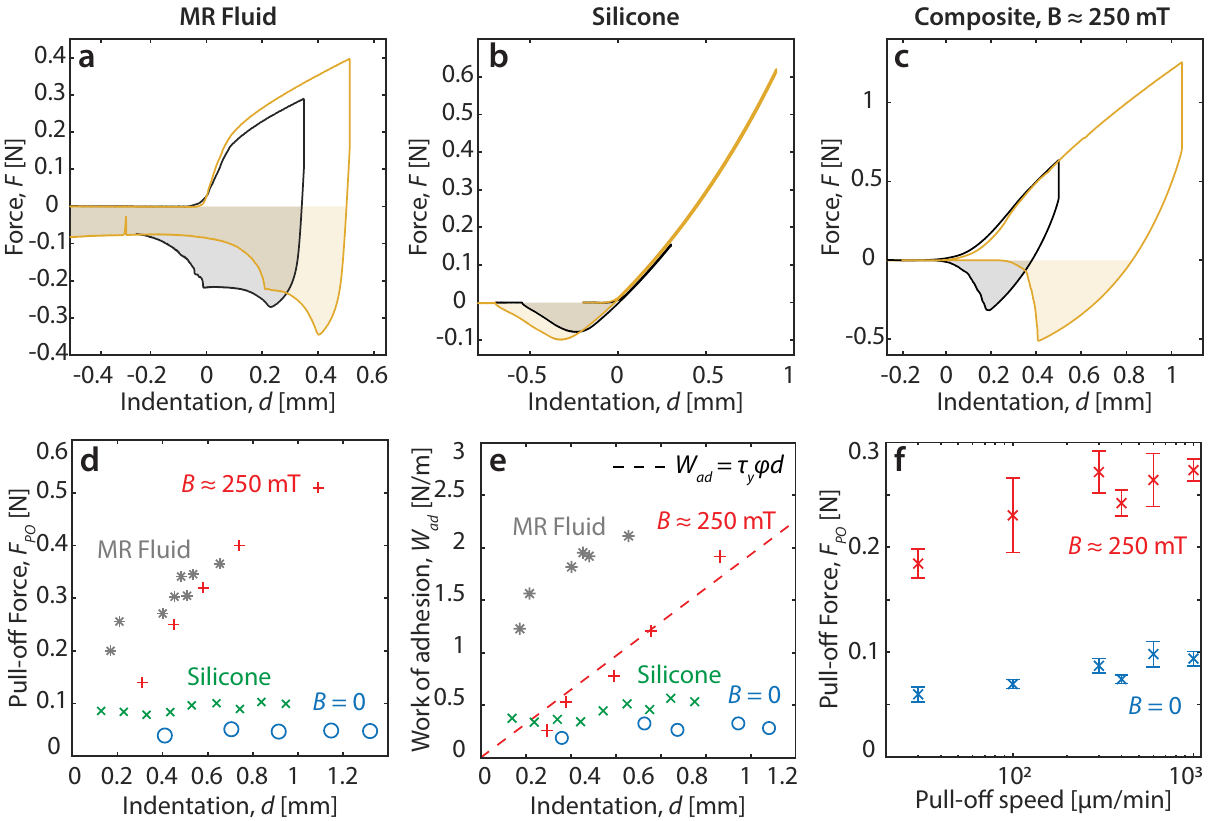}
\caption{\label{fig:indentation} 
Effect of indentation depth on adhesive strength.
a-c) Probe tack tests performed at two different indentation depths on a) pure MRF samples with $B\approx 250~\textrm{mT}$ , b) pure silicone with $E \approx 8.5$ kPa, and c) composite with $E \approx 12$ kPa and $B\approx 250~\textrm{mT}$.
d,e) Dependence of $F_{PO}$ and $W_{ad}$ on the maximum indentation depth for pure silicone, the pure magneto-rheological fluid and the composite without the applied magnetic field (blue points) and with (red points).
The dashed line corresponds to the scaling $W_{ad}=\tau_y \phi d$, with fitted  $\tau_y$=5.4 kPa.
f) Speed-dependence of the pull-off force for a $4~\mathrm{kPa}$ composite, with and without the magnetic field.
}
\end{figure*}

We quantified the contribution of dispersed MRF droplets to the dissipation  in a series of probe-tack tests.
Performing an indention cycle on a pure MRF in a magnetic field results in a large hysteresis loop (see Fig. \ref{fig:indentation}a).
If we indent further into the sample before retraction (yellow curve), there is more yielding, and even more energy dissipated, in both compression and tension. 
Deeper indentations also result in larger pull-off forces.
In contrast, for pure silicone, the total dissipated energy was unaffected by the total indentation depth, as shown in Fig. \ref{fig:indentation}b. 
For the composite, a dissipative behavior similar to that of the MRF is exhibited, which indicates that the MRF is responsible for this behaviour (Fig. \ref{fig:indentation}c).
To support this observation, the dependence of the pull-off force, $F_{PO}$, and work of adhesion, $W_{ad}$, on indentation depth is compared across materials in Fig. \ref{fig:indentation}d and \ref{fig:indentation}e, respectively.
When the magnetic field is off, the composite is insensitive to the indentation depth, which is similar to the behavior of the pure silicone.
When the magnetic field is on, the composite has higher, indentation-dependent, values of $W_{ad}$ and $F_{PO}$, mimicking the MRF.

This field-driven enhancement of adhesive properties  is very weakly rate dependent. 
The pull-off force, $F_{PO}$ is shown as a function of indentation speed in Fig. \ref{fig:indentation}f. 
Indeed, the $F_{PO}$ has a roughly logarithmic dependence on speed.
Since the magnetic field enhancement in the $F_{PO}$ persists even at very slow rates,
we conclude that the visco-elastic response of the composite is not the main driver of field-enhanced adhesion.
This supports the conclusion that the yield stress of the MRF is the  dominant rheological property contributing to field-enhanced adhesion.

We can estimate  the work of adhesion that can be achieved by this system, by considering the energy dissipated in the MRF inclusions during pull-off.
As the inclusions deform, they dissipate an energy per unit volume that scales with their yield stress, $\tau_y$.
Thus the total energy dissipated in the adhesive layer should scale like $W\sim\phi \tau_y d$, where $\phi$ is the volume fraction of MRF and $d$ is the indentation depth.
We fit the $B>0$ data in Fig. \ref{fig:indentation}e to this expression, while fixing the volume fraction to $\phi=0.3$ and varying the yield stress, $\tau_y$.  
This gives a reasonable fit, shown in Fig. \ref{fig:indentation}e, for a yield stress of  $ 5.4$ kPa, which is  consistent with measured value of the MRF yield stress at the applied field of $B \approx 250$ mT  (Fig. \ref{fig:overview}b), once again demonstrating that the MRF is responsible for the adhesive properties of the composite.

In conclusion, we have demonstrated magnetically-controlled adhesion of a composite elastomer to non-magnetic objects.
The composite is fabricated with a simple emulsion process, and performs well compared to previous magnetically-switchable adhesives (\emph{c.f.} Fig. \ref{fig:ashby-plot}).
The applied magnetic field increases the dissipation in the bulk of the composite, thereby changing the apparent work of adhesion and pull-off force.
The enhancement of the work of adhesion is captured by a simple model that accounts for the field-dependent yield stress of the inclusions, thus facilitating rational materials design.
The field-switchable adhesion is achievable at field strengths accessible with consumer permanent magnets as well as simple electromagnets, as demonstrated in  Movie S2.
These capabilities expand the utility of pressure sensitive adhesives to new applications requiring actuation of adhesive forces, such as soft robotic systems.

\section*{Materials \& Methods}
\textit{Sample preparation:} 
A water-based magneto-rheological fluid (formulated on request, Liquids Research Limited) with 80\% weight fraction of carbonyl iron microparticles was modified to avoid solvent evaporation by adding glycerol, resulting in a final fluid containing 66\% weight fraction of iron microparticles, 17\% weight fraction of glycerol, and 17\% weight fraction of water and stabilizers. 
At a relative humidity of $\approx$ 40\%, drying was not observed over the course of one week, as expected from the equilibrium composition of the water–glycerol mixture \cite{AssociationGlycerineProducers1953}. 
Silicone elastomers were produced by mixing different ratios of vinyl-terminated PDMS (DMS-V31, Gelest Inc.) with (25–35\% methylhydrosiloxane)-dimethylsiloxane copolymer, trimethylsiloxane terminated (HMS-301, Gelest Inc.), with the addition of platinum divinyl tetramethyldisiloxane catalyst (SIP6831.2, Gelest Inc.) according to the methodology reported by Style \emph{et al} \cite{Style2015} to obtain the desired elastic modulus.
To form the liquid precursor emulsion for the final composite, a 30\% volume fraction of the modified magneto-rheological fluid and the silicone precursors were mixed together with the surfactant molecule PEG-dimethicone (ES5612, DOW Corning) and stirred manually for 5 min. 
This resulted in an MRF-in-silicone emulsion, stabilized by the surfactant.
The mixture was degassed in a vacuum chamber for an additional 5 min, then poured into a $30 \times 30 $mm$^2$ wide  acrylic mold with a height of $h = 3$ mm and covered by a thin plastic sheet to obtain a flat and smooth surface, and left to crosslink overnight at room temperature.

\textit{Adhesion experiments:} 
Adhesion experiments were performed using a Zwick/Roell Z2.5 mechanical testing tool. 
The sample holders were clamped to the tool by an in-house built holder made of acrylic, in which a permanent magnet ($60\times60\times15$ mm$^3$, residual magnetic field = 1.3 T, Supermagnete) could be inserted to apply a magnetic field of $\approx 250$ mT. Indentation was performed using an acrylic cylindrical indenter of radius $a=4.8$ mm to avoid any artefacts in the force measurements in an applied magnetic field.
The test sequence consisted of an indentation step (speed = 1 mm/min), followed by a dwell time of 240 s, followed by a pull-off step (speed = 300 \textmugreek m/min).
During the dwell time we observed some viscoelastic relaxation of the force.

\textit{Adhesion experiments interpretation:} 
During indentation, the force increases linearly with displacement, as expected from linear elasticity.
The slope of this curve yields the Young's modulus $E$ given by
\begin{equation}
    E = \dfrac{F(1-\nu^2)}{2ad} g(a/h),
\end{equation}
where $F$ is the measured force, $d$ the indentation distance, $\nu$ is the Poisson's ratio of the polymer, assumed to be 0.5, and $a$ is the radius of the indenter.
$g$ is a correction factor that depends on the ratio between $a$ and the thickness of the sample, $h$, and is given by \cite{Lin2000}:
\begin{equation}
    g(a/h) = \left(1+\left(\dfrac{0.75}{(a/h)+(a/h)^3}+\dfrac{2.8(1-2\nu)}{(a/h)}\right)^{-1}\right)^{-1}
\end{equation}
We define the work of adhesion to be the integral of the force displacement curve in the tensile region ($F<0$) during pull-off divided by the area of the indenter:
\begin{equation}
    W_{ad} = \dfrac{\int_{ } F dd}{\pi a^2}
\end{equation}
The pull-off force, $F_{PO}$, is simply the maximum tensile force during pull-off.

\textit{Interface observations:}
In order to observe the interface between the indenter probe and the sample we inverted the indentation setup.
The sample and the magnet were then mounted on the moving arm of the mechanical testing tool, while the transparent acrylic indenter was fixed to a glass plate. 
A Thorlabs DCC3240M USB camera with a 0.5$\times$ to 1$\times$ telecentric objective (VariMagTL, Edmund Optics) was placed underneath the glass plate, and the indentation and retraction process was recorded.
The same test sequence that was used for adhesion experiments was applied.

\begin{figure}[h]
    \centering
    \includegraphics[width=8.5 cm]{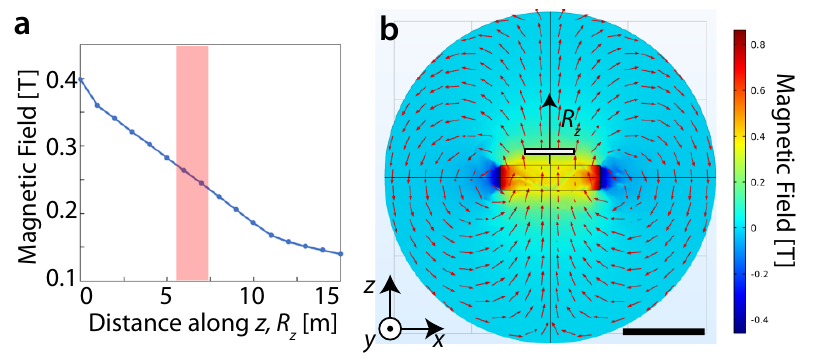}
    \caption{Simulation of magnetic field from the permanent magnet used in the study. 
    a) Out-of-plane component of the magnetic field as a function of the distance along $z$, labeled as $R_z$, from the magnet center.
    The magnetic field in the area of the measurement is indicated by the red shading, and has a magnitude of $\approx 250$ mT.
    b) Magnetic field intensity and direction in the $xz$ plane passing through the center of the magnet.
    The sample position is indicated by the white line.
    The arrow corresponds to the position of $Rz$, which is the line from which the data in (a) are extracted.
    The scale bar (bottom right) is 5 cm.}
    \label{fig:comsol-field}
\end{figure}

\textit{Magnetic field calculation:}
The magnetic field at the surface of the sample was determined using Comsol Multiphysics 5.4. For the magnetic field configuration shown in Fig. \ref{fig:comsol-field}, the field at the surface of the sample (distance along $z$ from the center of magnet of $\approx$ 6 mm) was found to be $\approx 250$ mT.

\begin{figure}
    \centering
    \includegraphics{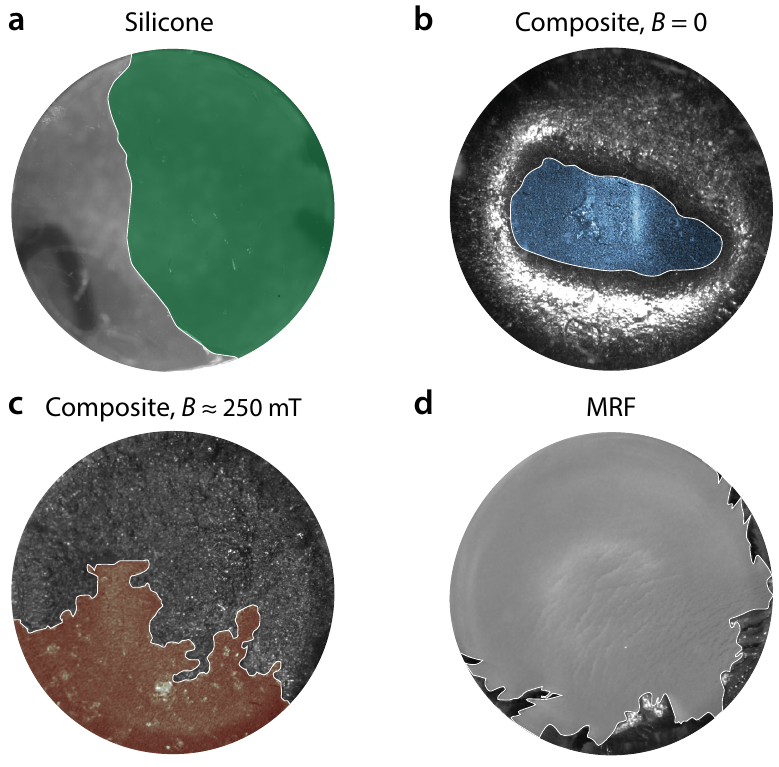}
    \caption{Images of the interfaces between a transparent indenter and different samples during  pull-off.
    a) Silicone, $E \approx 8.5$ kPa.
    b) Composite, $E \approx 4$ kPa and $B = 0~$.
    c) Composite, $E \approx 12$ kPa and $B\approx 250~\textrm{mT}$.
    d) Magneto-rheological fluid, $B\approx 250~\textrm{mT}$.}
    \label{fig:front-all}
\end{figure}

\textit{Indentation front comparison:}
Besides the snapshots shown in Fig. \ref{fig:crack_front}, a comparison of the indentation front with a pure silicone sample and a pure magneto-rheological fluid sample was performed using the same setup that was used for the indentation tests. 
The indentation front for different materials are shown in \ref{fig:front-all}, and for silicone (Fig. \ref{fig:front-all}a) is very smooth like that of the composite without magnetic field (Fig. \ref{fig:front-all}b), while the indentation front for the composite with magnetic field (Fig. \ref{fig:front-all}c) is rough, like that of the MRF (Fig. \ref{fig:front-all}d).
This supports the hypothesis that the shape of the indentation front is associated with the dissipation behavior at the interface, and ultimately to the values of $W_{ad}$ and $F_{PO}$.

\textit{Yield stress measurement:}
Magneto-rheological characterization of the MRF was performed with an Anton Paar MCR 302 rheometer equipped with a magneto-rheological device (MRD) measurement accessory (Anton Paar) in a 20 mm parallel plate configuration.
The accessory provides a magnetic field of up to 1000 mT during the measurement. 
The measurements were taken at 25$^\circ$ C.
Specifically, we performed a shear stress ramp measurement between $\tau = 0$ and $10^5 Pa$ at different fields between 0 and 1000 mT.
We extracted the yield stress from the $\tau$ vs $\dot{\gamma}$ plot as the linear interpolation of the slope after the onset of flow.

\section*{Conflicts of interest}
There are no conflicts to declare.

\section*{Acknowledgements}
The authors thank Chris Furrer for the help in creating the acrylic indenters, and Nicolas Bain, Qin Xu, Peter Derlet and Anand Jagota for helpful discussions.
The authors acknowledge the Paul Scherrer Institute, Villigen, Switzerland for provision of synchrotron radiation beamtime at the TOMCAT beamline X02DA of the SLS.
This work was funded by an ETH Research Grant (grant number ETH-48 17-1) ‘Tailored mesoscopic magneto-mechanical systems’, awarded for a project proposed by Paolo Testa, Laura J. Heyderman and Peter M. Derlet.

\section*{Author Contributions}
P.T., E.R.D., and L.J.H. developed the original material concept. R.W.S., E.R.D. and P.T. developed the adhesion measurements and interpretation. P.T., S.K. and B.C. designed and performed all the experiments with input from E.R.D. and R.W.S. P.T., R.W.S., and E.R.D. analyzed data. R.W.S. P.T., L.J.H., and E.R.D. wrote the paper.

\bibliography{library_SM}

\end{document}